\documentclass[11pt]{article} 
\usepackage{axodraw,mathrsfs}

\newcommand{\calE}{\mathcal E}
\newcommand{\calF}{\mathcal F}
\newcommand{\calG}{\mathcal G}
\newcommand{\calK}{K}
\newcommand{\calQ}{\mathcal Q}
\newcommand{\calP}{\mathcal P}
\newcommand{\submat}{\rm mat}
\newcommand{\calJ}{\mathscr J}

\title{
\centerline{\normalsize hep-ph/0006317 \hfill SINP/TNP/00/13}\bigskip
\bf Momentum-dependent contributions to the gravitational
coupling of neutrinos in a medium} 

\author{\bf Jos\'e F. Nieves\\
Laboratory of Theoretical Physics\\
Department of Physics, P.O. Box 23343 \\
University of Puerto Rico, R\'{\i}o Piedras \\
Puerto Rico 00931-3343\\[12pt]
\bf Palash B. Pal\\ 
Theory Group, Saha Institute of Nuclear Physics,\\ 1/AF Bidhan-Nagar, 
Calcutta 700064, India}
 
\date{}

\begin{document}
\maketitle

\begin{abstract}

When neutrinos travel through a normal matter medium, the electron
neutrinos couple differently to gravity compared to the other
neutrinos, due to the presence of electrons in the medium and the
absence of the other charged leptons.  We calculate the
momentum-dependent part of the matter-induced gravitational couplings
of the neutrinos under such conditions, which arise at order
$g^2/M^4_W$, and determine their contribution to the neutrino 
dispersion relation in the presence of a gravitational potential 
$\phi^{\mathrm{ext}}$.  
These new contributions vanish for the muon and tau neutrinos.
For electron neutrinos with momentum $K$, they are of the order of
the usual Wolfenstein term times the factor
$(K^2/M^2_W)\phi^{\mathrm{ext}}$, for high energy neutrinos.
In environments where the gravitational potential is substantial, such as those
in the vicinity of Active Galactic Nuclei,
they could be the dominant term in the neutrino dispersion
relation. They must also be taken into account in 
the analysis of possible violations of
the Equivalence Principle in the neutrino sector, in experimental
settings involving high energy neutrinos traveling through a 
matter background.

\end{abstract}

%
%
\section{Introduction}
\setcounter{equation}{0}
\label{sec:intro}
In the vacuum, a photon that propagates with momentum $\vec K$ in a
constant gravitational potential $\phi^{\rm ext}$ has a dispersion
relation given by
\begin{equation}
\label{photondisprel}
\omega_K = K(1 + 2\phi^{\rm ext}) \,.
\end{equation}
In fact, this formula holds not just for photons, but for any massless
particle. In particular, it also holds for neutrinos in the limit in
which their mass can be neglected.  This result is a consequence of
the universality of the gravitational interactions that is embodied in
the Principle of Equivalence, according to which gravity couples to
matter particles in a universal way; i.e., the graviton field couples
to the stress-energy tensor of the particles with a coupling constant
that is the same for all particle species.

Since the gravitational contribution to the neutrino dispersion
relation is the same for all the neutrino flavors, it is not relevant
in the context of neutrino oscillation experiments.  That contribution
yields a common factor in the evolution equation and therefore it
drops out of the formulas for the flavor transition amplitudes.

On the other hand, it has been pointed out by
Gasperini \cite{gasperini}, and by Halprin and Leung \cite{hl}, that a
small violation of the equivalence principle, manifest as a difference
in the couplings of the various neutrinos to gravity, would have
observable consequences in various neutrino experiments.  By the same
token, precise measurements of various observable quantities in these
experiments can place significant constraints on any deviations of the
Equivalence Principle in the neutrino sector\cite{constraints}.

In our earlier work \cite{gravnu,gressay} we considered neutrinos
traveling through a background of normal matter, made of electrons and
nucleons.  By calculating the weak interaction corrections to the
stress energy tensor of the neutrinos in the background, we showed
that the lowest order gravitational couplings of the neutrinos are
modified in a non-universal way.  In summary, the magnitude of the
background-induced contribution to the neutrino gravitational
couplings was calculated to order $g^2/M^2_W$, the corrections to the
neutrino dispersion relations in the presence of a gravitational
potential were determined, and some of their possible phenomenological
consequences were considered~\cite{agn}.

The corrections to the neutrino effective potential determined in
Ref.\ \cite{gravnu} are independent of the neutrino momentum because,
to order $g^2/M^2_W$, the weak interactions are local.  Our purpose in
the present work is to extend the calculation of the stress-energy
tensor of the neutrinos, to include the order $g^2/M^4_W$ terms. These
corrections arise by retaining the momentum-dependent terms of the
$W$-propagator in the relevant one-loop diagrams for the neutrino
gravitational vertex in matter.  The procedure is analogous to that
employed in Ref.\ \cite{nr,dnt} to determine the non-local corrections to
the neutrino effective potentials in matter in the absence of any
external field.

The distinguishing feature of the $O(g^2/M^4_W)$ corrections that we
determine in the present work is that they induce momentum-dependent
terms in the effective potential of the neutrinos in the presence of a
gravitational potential, which mimic the features of a non-universal
gravitational coupling of the neutrinos at a fundamental level.  We
point out that our calculations, and the results based on them, do not
depend on any physical assumption beyond those required by the
standard model of particle interactions and the linearized theory of
gravity, including the question of whether or not the neutrinos have a
non-zero mass.  Hence, the effects that we will consider are present
at some level and it is conceivable that they are detectable in some
favorable situations involving strong gravitational fields, such as
those that exist in the vicinity of active galactic nuclei.  In any
case, these corrections in principle must be taken into account in the
analysis of the tests and constraints of possible violations of the
Equivalence Principle in the neutrino sector, related to experimental
observations involving neutrinos traveling through a background, as
opposed to the vacuum.

The paper is organized as follows. In Sec.~\ref{sec:Frules}, we
present the diagrams to be calculated and the Feynman rules needed for
their calculation. We write down the expressions for vertex
diagrams with internal $Z$-boson and $W$-boson lines in
Sec.~\ref{sec:Z} and Sec.~\ref{sec:W} respectively. This sets up the
stage for Sec.~\ref{sec:selfenergy}, which contains most of the
results of this paper. In this section, we start by setting up the
general formalism which relates the neutrino gravitational vertex
function to the neutrino self-energy in an external gravitational
field. Then, in Sec.~\ref{subsec:gaugeinv}, we show that although the
general vertex can depend on the unphysical parameter $\xi$ appearing
in the $W$ and $Z$ propagators, the dispersion relation is independent
of it. The diagrams with internal $Z$-bosons lines do not contribute
to the dispersion relations at $O(g^2/M_W^4)$. The $O(g^2/M_W^4)$
contributions of the different $W$-mediated diagrams to the self
energy are calculated in Sec.~\ref{subsec:B4}, and they are
collected together in Sec.~\ref{subsec:disp} to find the neutrino
dispersion relation to $O(g^2/M_W^4)$. We conclude with some comments
in Section\ \ref{sec:conclusions}.

%
%
\section{Feynman rules}\label{sec:Frules}
\begin{figure}[thbp]
\begin{center}
%
%
\begin{picture}(100,170)(-50,-30)
\Text(0,-30)[cb]{\large\bf (A)}
\ArrowLine(40,0)(0,0)
\Text(35,-10)[cr]{$\nu_i(k)$}
\ArrowLine(0,0)(-40,0)
\Text(-35,-10)[cl]{$\nu_i(k^\prime)$}
\Photon(0,0)(0,40){2}{6}
\Text(-4,20)[r]{$Z$}
\ArrowArc(0,60)(20,90,270)
\ArrowArc(0,60)(20,-90,90)
\Text(23,60)[l]{$f(p)$}
\Text(-23,60)[r]{$f(p - q)$}
\Photon(0,80)(0,120){2}{6}
\Photon(0,80)(0,120){-2}{6}
\end{picture}
%
%
\begin{picture}(100,170)(-50,-30)
\Text(0,-30)[cb]{\large\bf (B)}
\ArrowLine(40,0)(0,0)
\Text(35,-10)[cr]{$\nu_i(k)$}
\ArrowLine(0,0)(-40,0)
\Text(-35,-10)[cl]{$\nu_i(k^\prime)$}
\Photon(0,0)(0,40){2}{6}
\Text(-4,20)[r]{$Z$}
\ArrowArc(0,60)(20,-90,270)
\Text(0,85)[b]{$f(p)$}
\Photon(0,20)(40,20){2}{6}
\Photon(0,20)(40,20){-2}{6}
\end{picture}
%
%
\begin{picture}(100,170)(-50,-30)
\Text(0,-30)[cb]{\large\bf (C)}
\ArrowLine(40,0)(0,0)
\Text(35,-10)[cr]{$\nu_i(k)$}
\ArrowLine(0,0)(-40,0)
\Text(-35,-10)[cl]{$\nu_i(k^\prime)$}
\Photon(0,0)(0,40){2}{6}
\Text(-4,20)[r]{$Z$}
\ArrowArc(0,60)(20,-90,270)
\Text(0,85)[b]{$f(p)$}
\Photon(0,0)(40,20){2}{6}
\Photon(0,0)(40,20){-2}{6}
\end{picture}
%
%
\begin{picture}(100,170)(-50,-30)
\Text(0,-30)[cb]{\large\bf (D)}
\ArrowLine(40,0)(0,0)
\Text(35,-10)[cr]{$\nu_i(k)$}
\ArrowLine(0,0)(-40,0)
\Text(-35,-10)[cl]{$\nu_i(k^\prime)$}
\Photon(0,0)(0,40){2}{6}
\Text(-4,20)[r]{$Z$}
\ArrowArc(0,60)(20,-90,270)
\Text(0,85)[b]{$f(p)$}
\Photon(0,40)(40,30){2}{6}
\Photon(0,40)(40,30){-2}{6}
\end{picture}
\end{center}

\caption[]{\sf $Z$-exchange diagrams for the one-loop contribution
to the gravitational vertex of any neutrino flavor
($i = e,\mu,\tau$) in a background of electrons and
nucleons. The braided line represents the graviton.
\label{fig:zdiags}
}
\end{figure}
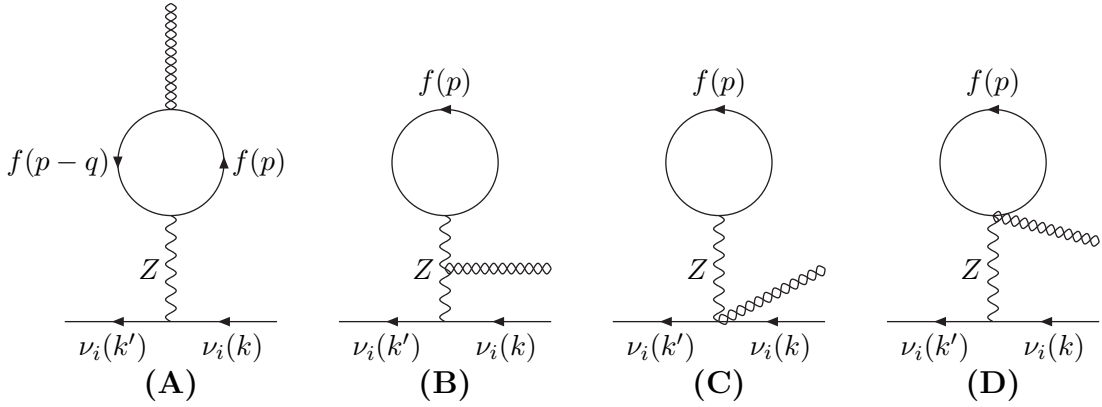

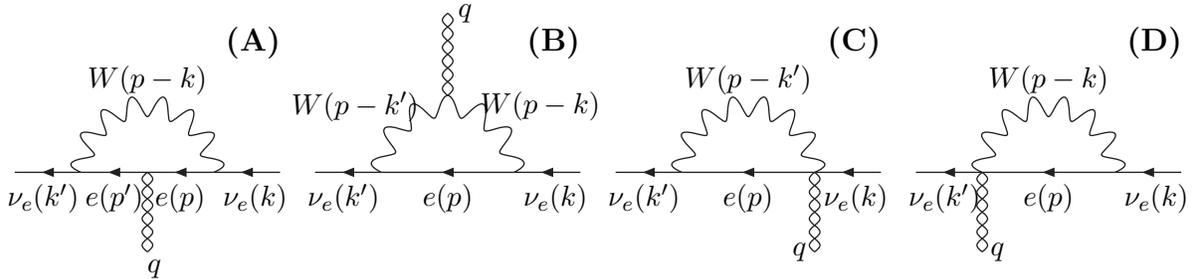
\begin{figure}[bhtp]
\begin{center}
%
%
\begin{picture}(110,100)(-55,-45)
\Text(40,55)[ct]{\large\bf (A)}
\ArrowLine(50,0)(25,0)
\Text(53,-10)[r]{$\nu_e(k)$}
\ArrowLine(25,0)(0,0)
\Text(12.5,-10)[c]{$e(p)$}
\ArrowLine(0,0)(-25,0)
\Text(-12.5,-10)[c]{$e(p')$}
\ArrowLine(-25,0)(-50,0)
\Text(-53,-10)[l]{$\nu_e(k')$}
\Photon(0,0)(0,-30){2}{3}
\Photon(0,0)(0,-30){-2}{3}
\Text(0,-33)[tl]{$q$}
\PhotonArc(0,0)(25,0,180){4}{7.5}
\Text(0,35)[c]{$W(p - k)$}
\end{picture}
%
%
\begin{picture}(110,100)(-55,-45)
\Text(40,55)[ct]{\large\bf (B)}
\ArrowLine(50,0)(25,0)
\Text(53,-10)[r]{$\nu_e(k)$}
\ArrowLine(25,0)(-25,0)
\Text(0,-10)[c]{$e(p)$}
\ArrowLine(-25,0)(-50,0)
\Text(-53,-10)[l]{$\nu_e(k')$}
\Photon(0,29)(0,60){2}{3}
\Photon(0,29)(0,60){-2}{3}
\Text(4,60)[l]{$q$}
\PhotonArc(0,0)(25,0,180){4}{6.5}
\Text(35,25)[c]{$W(p - k)$}
\Text(-35,25)[c]{$W(p - k')$}
\end{picture}
%
%
\begin{picture}(110,100)(-55,-45)
\Text(40,55)[ct]{\large\bf (C)}
\ArrowLine(50,0)(25,0)
\Text(53,-10)[r]{$\nu_e(k)$}
\ArrowLine(25,0)(-25,0)
\Text(0,-10)[c]{$e(p)$}
\ArrowLine(-25,0)(-50,0)
\Text(-53,-10)[l]{$\nu_e(k')$}
\PhotonArc(0,0)(25,0,180){4}{7.5}
\Text(0,35)[c]{$W(p - k')$}
\Photon(25,0)(25,-30){2}{3}
\Photon(25,0)(25,-30){-2}{3}
\Text(22,-30)[r]{$q$}
\end{picture}
%
%
%
\begin{picture}(110,100)(-55,-45)
\Text(40,55)[ct]{\large\bf (D)}
\ArrowLine(50,0)(25,0)
\Text(53,-10)[r]{$\nu_e(k)$}
\ArrowLine(25,0)(-25,0)
\Text(0,-10)[c]{$e(p)$}
\ArrowLine(-25,0)(-50,0)
\Text(-53,-10)[l]{$\nu_e(k')$}
\PhotonArc(0,0)(25,0,180){4}{7.5}
\Text(0,35)[c]{$W(p - k)$}
\Photon(-25,0)(-25,-30){2}{3}
\Photon(-25,0)(-25,-30){-2}{3}
\Text(-22,-30)[l]{$q$}
\end{picture}

\caption[]{
\sf $W$-exchange diagrams for the one-loop contribution
to the $\nu_e$ gravitational vertex in a background of electrons.
\label{fig:wdiags}
}
\end{center}
\end{figure}

\setcounter{equation}{0}
The diagrams that we must calculate in this paper are shown in Figs.\
\ref{fig:zdiags} and \ref{fig:wdiags}.  In principle, we must consider
also the diagrams in which any number of internal vector bosons is
replaced by the corresponding unphysical Higgs bosons. In general,
only by considering them we can attempt to prove that their
contribution cancels the one that arises from the gauge-dependent
piece of the gauge boson propagator.  However, for simplicity, we will
carry out the entire calculation in the limit $m_e\rightarrow 0$, in
which case the diagrams involving the unphysical Higgs particles do
not contribute. Hence, we do not consider them any further. Our task
is therefore twofold: to calculate the $O(1/M_B^4)$ contribution terms
of the diagrams shown in Figs.\ \ref{fig:zdiags} and \ref{fig:wdiags}
where $B$ is either the $W$ or the $Z$-boson, and to show that the
contribution from the gauge-dependent terms of the vector boson
propagators are zero in the limit $m_e\rightarrow 0$.
%
%
%
%

The propagators and vertices that we need to perform the computation
are as follows. The thermal propagator for a fermion of mass
$m_f$ is given by 
\begin{eqnarray}\label{SFf}
iS_f(p) = (\rlap/p + m_f) \left[
\frac{i}{p^2 - m_f^2 + i\epsilon} - 2\pi\delta(p^2 -
m_f^2)\eta_f(p) \right] \,, 
\end{eqnarray}
where
\begin{eqnarray}\label{etaf}
\eta_f(p) = \frac{\theta(p\cdot v)}{e^{\beta(p\cdot v - \mu_f)} + 1}
+ \frac{\theta(-p\cdot v)}{e^{-\beta(p\cdot v - \mu_f)} + 1}
\end{eqnarray}
with $\beta$ being the inverse temperature, $\mu_f$ the chemical
potential and $v^\mu$ the velocity four-vector of the medium.

The propagators for a gauge boson $B = W,Z$ are given in the $\xi$
gauge as
\begin{eqnarray}
\Delta_{B\mu\nu}(q) = \frac{-1}{q^2 - M_B^2}\left\{
\eta_{\mu\nu} - \frac{(1 - 1/\xi)q_\mu q_\nu}{q^2 - M_B^2/\xi}
\right\} \,.
\end{eqnarray}
The contributions of order $1/M_B^2$ to the neutrino effective 
gravitational vertex were calculated in Refs.\
\cite{gravnu,gressay}. 
Here we are interested in the terms next to the leading ones,
of order $1/M_B^4$. For that purpose, it is sufficient to 
approximate the boson propagators that appear in the loop integrals by
\label{propagatorapprox}
\begin{eqnarray}
\Delta_{B\mu\nu}(q) & = & \frac{\eta_{\mu\nu}}{M_B^2}
+ \Delta^{(4)}_{B\mu\nu}(q) + \Delta^{(\xi)}_{B\mu\nu}(q) \,,
\end{eqnarray}
where
\begin{eqnarray}
\Delta^{(4)}_{B\mu\nu}(q) & =  & \frac{1}{M_B^4}\left(
q^2\eta_{\mu\nu} - q_\mu q_\nu
\right) \,, \nonumber\\ 
\Delta^{(\xi)}_{B\mu\nu}(q) & = & \frac{\xi q_\mu q_\nu}{M_B^4} \,.
\end{eqnarray}
The remaining terms would yield contributions of order $1/M_B^6$ to
the effective vertex, which we are not considering.

%
%
\begin{figure}[thbp]
\begin{center}

\begin{picture}(180,60)(100,0)
\Photon(50,30)(100,0){-2}{7}
\ArrowLine(96,8)(86,14)
\Text(105,0)[l]{$B_\mu(k)$}
\Photon(50,30)(100,60){2}{7}
\ArrowLine(86,46)(96,52)
\Text(105,60)[l]{$B_\nu(k')$}
\Photon(0,30)(50,30){2}{7}
\Photon(0,30)(50,30){-2}{7}
\Text(25,35)[b]{$h_{\lambda\rho}$}
\Text(140,30)[l]{$= i\kappa M_B^2 a'_{\lambda\rho\mu\nu}
-i\kappa\left(C_{\lambda\rho\mu\nu}(k,k')
+ C^{(\xi)}_{\lambda\rho\mu\nu}(k,k')\right)$}
\end{picture}
\end{center}
\caption[]{
Feynman rule for the gravitational vertex coupling 
of a gauge boson $B = W,Z$ with the graviton. 
\label{fig:BBhvertex}
}
\end{figure} 
We also need the Feynman rules for the graviton couplings of the
vector bosons, including the dominant sub-leading terms
that were not considered in our earlier work, as well
as the gauge-dependent terms.
To obtain them we need to start from the free
Lagrangian, which for the $W$-bosons is
\begin{eqnarray}\label{LWg}
{\mathscr L}^{(W)}_g = \sqrt{\tt -g} \left\{ -\frac{1}{2}
g^{\mu\alpha} g^{\nu\beta} W^{\ast}_{\mu\nu}
W_{\alpha\beta} + M_W^2 g^{\mu\nu} W^\ast_\mu W_\nu
- \xi g^{\mu\nu}g^{\alpha\beta}(\partial_\mu W^\ast_\nu)
(\partial_\alpha W^\ast_\beta) \right\} \,,
\end{eqnarray}
where ${\tt g}= {\rm det}(g_{\mu\nu})$. We now write
\begin{eqnarray}
g_{\mu\nu} = \eta_{\mu\nu} + 2\kappa h_{\mu\nu} \,,
\end{eqnarray}
where $\kappa$ is related to the Newton's constant by
\begin{eqnarray}
\kappa = \sqrt{8\pi G} \,,
\end{eqnarray}
and $h_{\mu\nu}$ is the graviton field.  Keeping up to first order
terms in $\kappa$, we obtain the $WW$-graviton coupling in the form
\begin{eqnarray}
{\mathscr L}_h^{(WW)} = - \kappa h^{\lambda\rho}(x)
\left(
\widehat T_{\lambda\rho}(x) + \widehat T^{(\xi)}_{\lambda\rho}(x)
\right) \,,
\end{eqnarray}
where
\begin{eqnarray}
\widehat T_{\lambda\rho}(x) & = & \frac12 \eta_{\lambda\rho} W^*_{\mu\nu}
W^{\mu\nu} + 2 W^*_{\lambda\mu} W^\mu{}_\rho - M_W^2
a'_{\lambda\rho\mu\nu} W^*{}^\nu W^\mu \nonumber\\
\widehat T^{(\xi)}_{\lambda\rho}(x) & = & \xi\left\{
\eta_{\lambda\rho}(\partial\cdot W^\ast)(\partial\cdot W) -
(\partial_\lambda W^\ast_\rho + \partial_\rho W^\ast_\lambda)\partial\cdot W
- \partial\cdot W^\ast
(\partial_\lambda W_\rho + \partial_\rho W_\lambda)\right\} \,,
\end{eqnarray}
using the shorthand
\begin{eqnarray}
a'_{\lambda\rho\mu\nu} = \eta_{\lambda\rho} \eta_{\mu\nu} -
\big( \eta_{\lambda\mu} \eta_{\rho\nu} +
\eta_{\lambda\nu} \eta_{\rho\mu} \big) \,.
\label{a'}
\end{eqnarray}
The Feynman rule for the $WW$-graviton coupling can then be written as
in Fig.~\ref{fig:BBhvertex}, with $a'_{\lambda\rho\mu\nu}$ as in Eq.\
(\ref{a'}), and
\begin{eqnarray}
\label{C}
C_{\lambda\rho\mu\nu} (k,k') &=& \eta_{\lambda\rho} ( \eta_{\mu\nu} k
\cdot k' - k_\mu k'_\nu ) - \eta_{\mu\nu} (k_\lambda k'_\rho +
k'_\lambda k_\rho ) \nonumber\\*
&& \mbox{} + k_\nu (\eta_{\lambda\mu} k'_\rho +
\eta_{\rho\mu} k'_\lambda)
+ k'_\mu (\eta_{\lambda\nu} k_\rho +
\eta_{\rho\nu} k_\lambda) \nonumber\\*
&& \mbox{} - k \cdot k' (\eta_{\lambda\mu} \eta_{\rho\nu} +
\eta_{\lambda\nu} \eta_{\rho\mu}) \nonumber\\[12pt]
C^{(\xi)}_{\lambda\rho\mu\nu} (k,k') & = & \xi\left\{
\eta_{\lambda\rho}k_\mu k^\prime_\nu - k^\prime_\lambda k_\mu\eta_{\rho\nu}
- k^\prime_\rho k_\mu\eta_{\lambda\nu} - k_\lambda k^\prime_\nu\eta_{\rho\mu}
- k_\rho k^\prime_\nu \eta_{\lambda\mu}\right\} \,.
\end{eqnarray}
The rule for the couplings of the $Z$ boson are the same.

%
%
\begin{figure}[bhtp]
\begin{center}

\begin{picture}(180,70)(0,0)
\ArrowLine(0,0)(50,30)
\Text(20,55)[lb]{$h_{\lambda\rho}$}
\Text(80,55)[rb]{$Z_\alpha$}
\ArrowLine(50,30)(100,0)
\Photon(50,30)(100,60){2}{7}
\Photon(0,60)(50,30){2}{7}
\Photon(0,60)(50,30){-2}{7}
\Text(250,30)[r]{$= -i {g\over2\cos\theta_W} \kappa
a_{\lambda\rho\alpha\beta} \gamma^\beta (X_f+Y_f\gamma_5)$}
\end{picture}

\end{center}
\caption[]{
Feynman rule for the graviton coupling involving the 
$Z$-boson and a fermion. $a_{\lambda\rho\alpha\beta}$ has been defined 
in Eq.\ (\ref{defa}) and, in the standard model,
$X_\nu = -Y_\nu = 1/2$ for a neutrino, and
$-X_e = X_p = \frac{1}{2} - 2\sin^2\theta_W$,
$X_n =  -\frac{1}{2}$,
$Y_e = \frac{1}{2}$,
$Y_n = -Y_p = \frac{1}{2}g_A$,
with $g_A = 1.26$ being the renormalization constant
of the axial-vector current of the nucleon.
The analogous couplings involving the $W$ boson are
obtained by replacing ${g\over2\cos\theta_W}$ by ${g\over\surd2}$,
$X_f$ by $\frac12$ and $Y_f$ by $-\frac12$.
\label{fig:ffBh}
}
\end{figure}
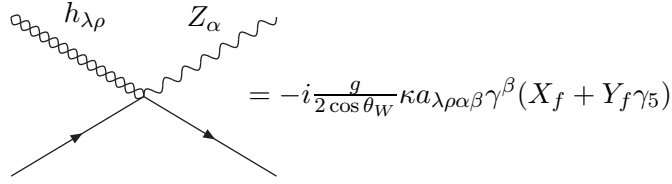 
The other couplings necessary for the present calculation
have been deduced in Ref.~\cite{gravnu}. 
For the gravitational coupling of a fermion of mass $m_f$,
the Feynman rule involves the factor
$-i\kappa V_{\lambda\rho}(p,p')$, with
\begin{eqnarray}
V_{\lambda\rho}(p,p') = \frac14 \Big[ \gamma_\lambda (p+p')_\rho +
\gamma_\rho (p+p')_\lambda \Big] - \frac12 \Big[ \rlap/p + \rlap/p' -
2m_f \Big] \,,
\end{eqnarray}
where $p$ and $p'$ denote
the momentum of the incoming and the outgoing fermion, respectively.
There is also a fermion vertex involving
both a graviton and a gauge boson line, and its Feynman rule is
given in Fig.~\ref{fig:ffBh}, with the definition
\begin{eqnarray}
a_{\lambda\rho\alpha\beta} 
& = & \eta_{\lambda\rho}\eta_{\alpha\beta} - \frac{1}{2}(
\eta_{\lambda\alpha}\eta_{\rho\beta} + 
\eta_{\rho\alpha}\eta_{\lambda\beta}) \,.
\label{defa} 
\end{eqnarray}
%

%
%
\section{Calculation of the vertex function diagrams}
\setcounter{equation}{0}
\subsection{The $Z$-mediated diagrams}\label{sec:Z}
The diagrams, which are shown in Fig.~\ref{fig:zdiags}, are the same as
in Ref.~\cite{gravnu}. We use the same notation and conventions given
in that reference for the weak-neutral-current couplings in the Lagrangian 
[e.g., Eqs.\ (2.31)-(2.35) there], summarized in Fig.~\ref{fig:ffBh} here.
The $Z$-propagator is external to the loop, so
all the algebraic manipulations of the loop integral are performed 
exactly as in Ref.~\cite{gravnu}. As in that reference, we define
\begin{eqnarray}\label{bZ}
b^{(Z)} = \sum_{f} X_f b_f \,,
\end{eqnarray}
where the sum is over the fermions species in the background, 
and\footnote{
We take the opportunity to correct a typographical
error in Eq. (3.20) of Ref.\ \cite{gravnu}
}
\begin{eqnarray}\label{b}
b_f & = & 4\sqrt{2}G_F
\int\frac{d^4p}{(2\pi)^3}\delta(p^2- m_f^2) \eta_f(p) \; p\cdot v\nonumber\\
& = & \sqrt{2}G_F(n_f - n_{\bar f}) \,, 
\end{eqnarray}
with $n_{f,{\bar f}}$ being the number densities of the particles
and antiparticles respectively.
We denote by $k,k'$ the momentum of the initial and final neutrino 
respectively, and the momentum of the outgoing graviton is
\begin{eqnarray}
q = k - k^\prime \,.
\end{eqnarray}
For Fig.~\ref{fig:zdiags}A we then obtain
\begin{eqnarray}
\Gamma'{}^{(\ref{fig:zdiags}A)}_{\lambda\rho} =
\left( \gamma^\alpha + {q^2 \gamma^\alpha -
(1 - \xi)q^\alpha \rlap/q \over M_Z^2} \right)\left(
\Lambda^{(Z)}_{\lambda\rho\alpha} - 
b^{(Z)} \eta_{\lambda\rho} v_\alpha\right)  L \,,
\label{GamZAfinal}
\end{eqnarray}
where
\begin{eqnarray}\label{LambdaZfinal}
\Lambda^{(Z)}_{\lambda\rho\alpha} &=& {g_Z^2 \over M_Z^2}
\sum_{f}\int\frac{d^3p}{2E_f(2\pi)^3}
\left\{X_f(f_f - f_{\overline f}) \left[
\frac{N^{(1)}_{\lambda\rho\alpha}(p_f,q)}{q^2 - 2p_f\cdot q} + 
(q\rightarrow -q)\right] \right. \nonumber\\
& & \mbox{} \left.
- Y_f(f_f + f_{\overline f})\left[
\frac{N^{(2)}_{\lambda\rho\alpha}(p_f,q)}{q^2 - 2p_f\cdot q} -
(q\rightarrow -q) \right] \right\} \,,
\end{eqnarray}
with $f_f$ and $f_{\overline f}$ denoting the momentum
distribution functions for
particles and antiparticles and
\begin{eqnarray}
N^{(1)}_{\lambda\rho\alpha}(p,q) & \equiv &
(2p - q)_\lambda \big[2p_\rho p_\alpha
- (p_\alpha q_\rho + q_\alpha p_\rho) + (p\cdot q)\eta_{\alpha\rho}
\big] 
+ (\lambda\leftrightarrow\rho) \nonumber\\
N^{(2)}_{\lambda\rho\alpha}(p,q) & \equiv &
(2p - q)_\lambda i\epsilon_{\rho\alpha\beta\sigma}q^\beta p^\sigma +
(\lambda\leftrightarrow\rho) \,.
\label{N}
\end{eqnarray}

In Fig.~\ref{fig:zdiags}B, the $Z$ line attached to the fermion loop
carries zero momentum, in which case the $C_{\lambda\rho\mu\nu}$
and $C^{(\xi)}_{\lambda\rho\mu\nu}$ terms of the $ZZ$ graviton couplings
do not contribute. Hence the contribution from this diagram is given by
\begin{eqnarray}
\Gamma^{(\ref{fig:zdiags}B)}_{\lambda\rho} & = & -b^{(Z)}
a^\prime_{\lambda\rho\alpha\beta} \left( \gamma^\alpha + {q^2
\gamma^\alpha - 
(1 - \xi)q^\alpha \rlap/q \over M_Z^2} \right) L v^\beta \,,
\label{GamZBfinal}
\end{eqnarray}
where $a'_{\lambda\rho\alpha\beta}$ has been defined in Eq.\ (\ref{a'}).

In Fig.~\ref{fig:zdiags}C, the $Z$-boson line has zero momentum
flowing through it. As a consequence, the $O(M_Z^{-4})$
terms of the $Z$-propagator do not contribute and 
the amplitude for this diagram is the same as obtained 
in Refs.~\cite{gravnu,gressay}, viz.,
\begin{eqnarray}
\Gamma^{(\ref{fig:zdiags}C)}_{\lambda\rho} & = & b^{(Z)}
a_{\lambda\rho\alpha\beta} \gamma^\alpha L v^\beta \,,
\label{GamZCfinal}
\end{eqnarray}
where $a_{\lambda\rho\alpha\beta}$ was defined in Eq.\ (\ref{defa}).
Finally, for Fig.~\ref{fig:zdiags}D, we get 
\begin{eqnarray}
\Gamma^{(\ref{fig:zdiags}D)}_{\lambda\rho} & = & b^{(Z)}
a_{\lambda\rho\alpha\beta} \left( \gamma^\alpha + {q^2 \gamma^\alpha -
(1 - \xi)q^\alpha \rlap/q \over M_Z^2} \right) L v^\beta \,.
\label{GamZDfinal}
\end{eqnarray}

In this way, summing up all the diagrams 
of Fig.~\ref{fig:zdiags}, we obtain the total
contribution from the $Z$-diagrams to the effective vertex 
\begin{eqnarray}\label{zexchange}
\Gamma'{}^{(\ref{fig:zdiags})}_{\lambda\rho} & = &
\Lambda^{(Z)}_{\mu\nu\alpha}\gamma^\alpha L \nonumber\\
& & +
\left( 
{q^2 \gamma^\alpha - (1 - \xi)q^\alpha \rlap/q \over M_Z^2} \right)L\left(
\Lambda^{(Z)}_{\mu\nu\alpha} + b^{(Z)} \Big[
\eta_{\lambda\rho}\eta_{\alpha\beta} + 
\frac12 ( \eta_{\lambda\alpha}\eta_{\rho\beta} + 
\eta_{\rho\alpha}\eta_{\lambda\beta}) \Big]  
v^\beta \right)\,,
\end{eqnarray}
where the first term on the right hand side represents the leading
order contribution, of order $(1/M_Z^2)$, whose detailed evaluation
was the subject of Ref. \cite{gravnu}.  It was shown in that reference
that, for a static and constant gravitational potential, which
involves taking the $q\rightarrow 0$ of the vertex function in the
proper way, the contribution from the $\Lambda^{(Z)}_{\mu\nu\alpha}$
term to the neutrino self-energy is a well defined quantity.
Borrowing that result here, it follows that the remaining terms in
Eq.\ (\ref{zexchange}), which involve the same quantity
$\Lambda^{(Z)}_{\mu\nu\alpha}$, yield a contribution to the neutrino
self-energy that depends on the derivatives of
the gravitational field. Therefore, in the presence of a static
and homogeneous potential, the $O(1/M_Z^4)$ contributions to the neutrino
self-energy vanish. Since these are the only diagrams that contribute
to the $\nu_{\mu,\tau}$ dispersion relation if there is no $\mu$ or
$\tau$ in the background, as we assume, only the dispersion relation
of the $\nu_e$ is modified by the $O(1/M^4_B)$ terms. These arise from
the $W$ diagrams that we consider next.

\subsection{The $W$-mediated diagrams}\label{sec:W}
We now discuss the diagrams involving the $W$-bosons, shown in
Fig.~\ref{fig:wdiags}, which are relevant only for the
electron neutrino. In what follows, unless we explicitly mention it
otherwise, we are referring only to this type of neutrino. 

\subsubsection{Diagram \ref{fig:wdiags}A}
Let us start with Fig.~\ref{fig:wdiags}A. It gives
\begin{eqnarray}
\Gamma^{(\ref{fig:wdiags}A)}_{\lambda\rho} = - \frac{ig^2}{2}
\int\frac{d^4p}{(2\pi)^4} \gamma^\alpha L iS_e(p') 
V_{\lambda\rho}(p,p') iS_e(p)\gamma^\beta L \Delta_{W\alpha\beta}(p -
k)\,, 
\label{GamWA}
\end{eqnarray}
where
\begin{eqnarray}
p^\prime = p - q\,.
\end{eqnarray}
%
%
%
By substituting the $W$ propagator in Eq.\ (\ref{GamWA})
three terms are produced. The $O(1/M_W^2)$ term
has already been calculated in Ref.~\cite{gravnu}. 
Separating it out, we can write the dispersive part 
of $\Gamma'{}^{(\ref{fig:wdiags}A)}_{\lambda\rho}$ in the form
\begin{eqnarray}\label{defGA}
\Gamma'{}^{(\ref{fig:wdiags}A)}_{\lambda\rho} &=&
\Lambda^{(W)}_{\lambda\rho} - 
b_e\rlap/ v L\eta_{\lambda\rho} +
G^{(A)}_{\lambda\rho} + H^{(A)}_{\lambda\rho} \,,
\end{eqnarray}
where $G^{(A)}_{\lambda\rho}$, which contains all the $O(1/M_W^4)$
contributions that are $\xi$-independent, is given by
\begin{eqnarray}
\label{GA}
G^{(A)}_{\lambda\rho} &=& - \frac{g^2}{2}
\int\frac{d^4p}{(2\pi)^3} \gamma^\alpha L \rlap/p' 
V_{\lambda\rho}(p,p') \rlap/p 
\gamma^\beta L \; \Delta^{(4)}_{W\alpha\beta}(p - k)
%
%
\nonumber\\*
&\times& \left( {\delta(p^2) \eta_e(p)
\over p'^2} + {\delta(p'^2) \eta_e(p')
\over p^2} \right) \,,
\end{eqnarray}
while $H^{(A)}_{\lambda\rho}$ is given by the same expression 
as Eq.\ (\ref{GA}) but with the replacement
\begin{eqnarray}
\label{replacement} 
\Delta^{(4)}_{W\alpha\beta} \rightarrow \Delta^{(\xi)}_{W\alpha\beta} \,.
\end{eqnarray}
%

\subsubsection{Diagram \ref{fig:wdiags}B}
Proceeding in similar fashion
for this diagram, we borrow the leading order result derived
in Ref.~\cite{gravnu} and write
\begin{eqnarray}
\Gamma'{}^{(\ref{fig:wdiags}B)}_{\lambda\rho} = 
b_e \Big[-\eta_{\lambda\rho}\rlap/ v - \frac{1}{2} 
(\gamma_\lambda\rlap/ v\gamma_\rho + \gamma_\rho\rlap/
v\gamma_\lambda) \Big]L + G^{(B)}_{\lambda\rho} +
H^{(B)}_{\lambda\rho}\,, 
\label{defGB}
\end{eqnarray}
where $G^{(B)}_{\lambda\rho}$ and $H^{(B)}_{\lambda\rho}$ contain the
non-leading terms, separated according to whether they are
$\xi$\-independent or not, respectively. These non-leading terms can
come either from the non-leading contribution in the $WW$-graviton
coupling or in the $W$-propagator. Accordingly, we get the following
terms
\begin{eqnarray}
\label{GHB}
G^{(B)}_{\lambda\rho} &=& \frac{g^2}{2}
\int\frac{d^4p}{(2\pi)^3}\delta(p^2)\eta_e(p)
\gamma^\alpha L\rlap/p 
\gamma^\beta L \nonumber\\* 
&\times& 
\Bigg[-\frac{1}{M_W^4}C_{\lambda\rho\alpha\beta}(p - k^\prime,p - k)  
+ \eta^{\mu\nu} a^\prime_{\lambda\rho\alpha\mu}
\Delta^{(4)}_{W\nu\beta}(p - k) 
+ \eta^{\mu\nu} a^\prime_{\lambda\rho\mu\beta}
\Delta^{(4)}_{W\alpha\nu}(p - k')
\Bigg]\nonumber\\[12pt]
H^{(B)}_{\lambda\rho} &=& \frac{g^2}{2}
\int\frac{d^4p}{(2\pi)^3}\delta(p^2)\eta_e(p)
\gamma^\alpha L \rlap/p \gamma^\beta L\nonumber\\
& \times &
\Bigg[
- \frac{1}{M^4_W}C^{(\xi)}_{\lambda\rho\alpha\beta}(p - k^\prime,p - k)
+ \eta^{\mu\nu} a^\prime_{\lambda\rho\alpha\mu}
\Delta^{(\xi)}_{W\nu\beta}(p - k) 
+ \eta^{\mu\nu}a^\prime_{\lambda\rho\mu\beta}
\Delta^{(\xi)}_{W\alpha\nu}(p - k')
\Bigg] \nonumber\\
\end{eqnarray}
%

\subsubsection{Diagrams \ref{fig:wdiags}C and D}
Similarly, we define the non-leading contributions by writing
\begin{eqnarray}
\label{defGC}
\Gamma^{(\ref{fig:wdiags}C)}_{\lambda\rho} & = &
b_e \Big[\eta_{\lambda\rho}\rlap/ v + \frac14
(\gamma_\lambda\rlap/ v\gamma_\rho + \gamma_\rho\rlap/
v\gamma_\lambda) \Big]L + G^{(C)}_{\lambda\rho} + H^{(C)}_{\lambda\rho}
 \,, \\*
\label{defGD}
\Gamma^{(\ref{fig:wdiags}D)}_{\lambda\rho} & = & 
b_e \Big[\eta_{\lambda\rho}\rlap/ v + \frac14
(\gamma_\lambda\rlap/ v\gamma_\rho + \gamma_\rho\rlap/
v\gamma_\lambda) \Big]L + G^{(D)}_{\lambda\rho} +
H^{(D)}_{\lambda\rho}\,.
\end{eqnarray}
For these diagrams, the non-leading terms come only from the $W$
propagator. Thus we obtain 
\begin{eqnarray}
G^{(C)}_{\lambda\rho} &=& -\, {g^2 \over 2} 
a_{\lambda\rho\beta\mu}
\int\frac{d^4p}{(2\pi)^3}\delta(p^2)\eta_e(p)
\gamma_\alpha L \rlap/p \gamma^\beta L
\Delta^{(4)\mu\alpha}_W(p - k') 
\label{GC}\\*
G^{(D)}_{\lambda\rho} &=& -\, {g^2 \over 2} 
a_{\lambda\rho\alpha\mu}
\int\frac{d^4p}{(2\pi)^3}\delta(p^2)\eta_e(p)
\gamma^\alpha L \rlap/p \gamma_\beta L 
\Delta^{(4)\beta\mu}_W(p- k) 
\label{GD}
\end{eqnarray}
The corresponding formulas for $H^{(C,D)}_{\lambda\rho}$ are given by
the same expressions, with the replacement indicated in Eq.\
(\ref{replacement}).

%
%
\section{Self-energy in a constant potential}\label{sec:selfenergy}
\setcounter{equation}{0}
\subsection{General formalism}
%
In the presence of a static and homogeneous gravitational potential,
the neutrino self-energy, including the $O(g^2/M_W^4)$ corrections,
is given by
\begin{eqnarray}
\label{Sigmatot}
\Sigma = \Sigma_{\submat} + 
\Sigma^{(2)}_G + \Sigma^{(4)}_G + \Sigma^{(\xi)}_G
\end{eqnarray}
where $\Sigma_{\submat}$ stands for the purely weak self-energy terms
which do not depend on the gravitational potential, including
the $O(g^2/M^4_W)$ corrections, while
$\Sigma^{(2)}_G$ is the quantity calculated in Ref.\ \cite{gravnu}
[defined in Eq.\ (4.8) there].  
In analogy with the
expression given in Eq. (4.9) of that reference, we can write here
\begin{eqnarray}
\label{sigma4}
\Sigma^{(4)}_G(\omega,\vec K) & = & \phi^{\rm ext} (2 v^\lambda v^\rho -
\eta^{\lambda\rho}) G_{\lambda\rho} (0,\vec{\cal Q}\rightarrow 0) \,,
\nonumber\\ 
\Sigma^{(\xi)}_G(\omega,\vec K) & = & \phi^{\rm ext} 
(2 v^\lambda v^\rho -\eta^{\lambda\rho}) 
H_{\lambda\rho}(0,\vec{\cal Q}\rightarrow 0) \,,
\end{eqnarray}
where $G_{\lambda\rho}(\Omega,\vec\calQ)$ and
$H_{\lambda\rho}(\Omega,\vec\calQ)$ denote the sum of the various
gauge-independent and gauge-dependent terms $G^{(X)}_{\lambda\rho}$
and $H^{(X)}_{\lambda\rho}$ respectively, whose integral expressions
have been given in Eqs.\ (\ref{GA}), (\ref{GHB}), (\ref{GC}) and
(\ref{GD}).  As we already argued at the end of Sec.~\ref{sec:Z}, the
$O(g^2/M_Z^4)$ terms from the $Z$-mediated diagrams do not contribute
to the self-energy of the neutrinos, and we have to calculate the
contributions only from the $W$-mediated diagrams in Eq.\
(\ref{sigma4}).  In Eq.\ (\ref{sigma4}) we have indicated
the dependence of the various quantities\footnote{Although we do not
show it explicitly, $G^{(X)}$ depends on $\omega$ and $K$ as well.} 
on the kinematic variables of the problem. These variables are 
defined by writing 
\begin{eqnarray}
\label{kqatrest}
k^\mu & = & (\omega,\vec K) \nonumber\\
q^\mu & = & (\Omega,\vec\calQ) 
\end{eqnarray}
in the rest frame of the medium where
\begin{equation}
\label{vatrest}
v^\mu = (1,\vec 0) \,.
\end{equation}
In what follows, we indicate the dependence of the various functions
on these variables when it is necessary, but we omit it otherwise. 

The chirality of the neutrino interactions dictate that $\Sigma$ has
the decomposition
\begin{eqnarray}
\label{chirality}
\Sigma = (a k\llap / + b v\llap /)L \,,
\end{eqnarray}
where $a$ and $b$ are in general functions of $\omega$ and $K$.  It is
convenient to split $a$ and $b$ as
\begin{eqnarray}
a & = & a_{\submat} + a^{(2)}_G + a^{(4)}_G + a^{(\xi)}_G\nonumber\\
b & = & b_{\submat} + b^{(2)}_G + b^{(4)}_G + b^{(\xi)}_G \,,
\end{eqnarray}
corresponding to the decomposition given in Eq.\ (\ref{Sigmatot}).
%
%
Eq.\ (\ref{chirality}) implies that $a^{(4)}_G$ and $b^{(4)}_G$ can be
determined by means of the formulas
\begin{eqnarray}
\label{abABrel}
a^{(4)}_G & = & \frac{\omega A^{(4)} - B^{(4)}}{K^2}
\nonumber\\
b^{(4)}_G & = & \frac{\omega B^{(4)} - k^2 A^{(4)}}{K^2}\,, 
\end{eqnarray}
and similarly
\begin{eqnarray}
a^{(\xi)}_G & = & \frac{\omega A_\xi - B_\xi}{K^2}\nonumber\\
b^{(\xi)}_G & = & \frac{\omega B_\xi - k^2 A_\xi}{K^2} \,,
\end{eqnarray}
where
\begin{eqnarray}
\label{AB4}
B^{(4)}(\omega,K) & = & \phi^{\rm ext}
(2 v^\lambda v^\rho -\eta^{\lambda\rho}) 
\frac{1}{2}\mbox{Tr}k\llap / 
G_{\lambda\rho}(0,{\cal Q}\rightarrow 0) \,,
\nonumber\\ 
A^{(4)}(\omega,K) & = & \phi^{\rm ext} 
(2 v^\lambda v^\rho -\eta^{\lambda\rho}) 
\frac{1}{2}\mbox{Tr}v\llap / 
G_{\lambda\rho}(0,{\cal Q}\rightarrow 0) \,,
\end{eqnarray}
and similarly
\begin{eqnarray}
\label{ABxi}
B_\xi(\omega,K) & = & \phi^{\rm ext}
(2 v^\lambda v^\rho -\eta^{\lambda\rho}) 
\frac{1}{2}\mbox{Tr}k\llap / 
H_{\lambda\rho}(0,{\cal Q}\rightarrow 0) \,, \nonumber\\
A_\xi(\omega,K) & = & \phi^{\rm ext}
(2 v^\lambda v^\rho -\eta^{\lambda\rho}) 
\frac{1}{2}\mbox{Tr}v\llap / 
H_{\lambda\rho} (0,{\cal Q}\rightarrow 0) \,.
\end{eqnarray}
As discussed in Ref.\ \cite{gravnu}, the dispersion relations
are obtained by solving
\begin{eqnarray}
\omega_K & = & K + \frac{b(\omega_K,K)}{1 - a(\omega_K,K)}\nonumber\\
\bar\omega_K & = & K - \frac{b(-\bar\omega_K,K)}{1 - a(-\bar\omega_K,K)}
\end{eqnarray}
for the neutrinos and antineutrinos, respectively. In the context
of our perturbative approach, the solutions are given
approximately by 
\begin{eqnarray}
\omega_K & = & K + [1 + a(K,K)]b(K,K) \nonumber\\
\bar\omega_K & = & K - [1 + a(-K,K)]b(-K,K) \,.
\end{eqnarray}

Using the result for 
the dispersion relation obtained in Ref.\ \cite{gravnu}
[given in Eq. (4.32) of that reference], it is easy to see
from Eq.\ (\ref{abABrel}) that in the present case
\begin{eqnarray}
\omega_K & = & K + 2\phi^{\rm ext}K + (1 + \phi^{\rm ext})b_{\submat} + 
b^{(2)}_G +
\frac{B^{(4)}(K,K)}{K} + 
\frac{B_\xi(K,K)}{K} \nonumber\\
\bar\omega_K & = & K + 2\phi^{\rm ext}K + 
(1 + \phi^{\rm ext}){\bar b}_{\submat} - b^{(2)}_G +
\frac{B^{(4)}(-K,K)}{K} + \frac{B_\xi(-K,K)}{K} \,.
\end{eqnarray}
Here $b^{(2)}_G$ is the contribution calculated in Ref.\ \cite{gravnu}
[denoted by simply $b_G$ and given by the formula in Eq.\ (4.30)
there], while $b_{\submat}$ and ${\bar b}_{\submat}$ are the usual
contributions independent of the gravitational potential, for
neutrinos and antineutrinos respectively, including the $O(g^2/M^4_W)$
terms. These can be obtained from the references cited\cite{nr,dnt},
and we will quote explicit formulas for our cases of interest in
Sec.~\ref{subsec:disp}.  Thus, what remains is simply to calculate the
coefficient $B^{(4)}$ by means of Eq.\ (\ref{AB4}), using the integral
formulas for the various functions $G^{(X)}_{\lambda\rho}$ obtained in
the previous section.  However, before we proceed to do that, we will
now show that the quantity $B_\xi(\pm K,K)$, which is calculated in
similar fashion, is in fact zero so that we finally obtain the gauge
invariant result
\begin{eqnarray}
\label{disprelformula}
\omega_K & = & K + 2\phi^{\rm ext}K + (1 + \phi^{\rm ext})b_{\submat} + 
b^{(2)}_G +
\frac{B^{(4)}(K,K)}{K} \nonumber\\
\bar\omega_K & = & K + 2\phi^{\rm ext}K + 
(1 + \phi^{\rm ext}){\bar b}_{\submat} - b^{(2)}_G +
\frac{B^{(4)}(-K,K)}{K} \,.
\end{eqnarray}
%

\subsection{Gauge independence of dispersion
relations}\label{subsec:gaugeinv} 
We calculate $B_\xi$ using Eq.\ (\ref{ABxi}).
We denote by $B^{(X)}_\xi$ the contribution from each term
$H^{(X)}_{\lambda\rho}$ that arises in that formula.  Carrying
out the trace operation first, a straightforward calculation yields
the expressions
\begin{eqnarray}
\label{Bxiresults}
B^{(A)}_\xi(\omega,K) & = & \phi^{\rm ext}\left(-\frac{g^2\xi}{2
M_W^4}\right)k^2 
(2 v^\lambda v^\rho - \eta^{\lambda\rho})
\left[
X^{(1)}_{\lambda\rho} + Y_{\lambda\rho}(0,{\cal Q}\rightarrow 0)
\right] \nonumber\\
B^{(B)}_\xi(\omega,K) & = & \phi^{\rm ext} \left( -\frac{g^2\xi}{2
M_W^4}\right)k^2 
(2 v^\lambda v^\rho - \eta^{\lambda\rho})
\left[
a^\prime_{\lambda\rho\alpha\beta} X^{(2)\alpha\beta} + X^{(3)}_{\lambda\rho} 
\right]\nonumber\\
B^{(C)}_\xi(\omega,K) + B^{(D)}_\xi(\omega,K) & = & 
\phi^{\rm ext}
\left(\frac{g^2\xi}{2 M_W^4}\right)k^2 
(2 v^\lambda v^\rho - \eta^{\lambda\rho})
a_{\lambda\rho\alpha\beta}
X^{(2)\alpha\beta} \,,
\end{eqnarray}
where
\begin{eqnarray}
\label{I12J}
X^{(1)}_{\lambda\rho} & = &
-\int\frac{d^4p}{(2\pi)^3}\delta(p^2)\eta_e(p)\left[
\frac{1}{2}\eta_{\lambda\rho}(k\cdot p) + 2p_\lambda p_\rho\right] 
\nonumber\\
X^{(2)}_{\alpha\beta} & = &
\int\frac{d^4p}{(2\pi)^3}\delta(p^2)\eta_e(p)\left[
2p_\alpha p_\beta - p_\alpha k_\beta - p_\beta k_\alpha\right]
\nonumber\\ 
X^{(3)}_{\lambda\rho} & = &
\int\frac{d^4p}{(2\pi)^3}\delta(p^2)\eta_e(p)\left[
\eta_{\lambda\rho}p\cdot k + 4p_\lambda p_\rho 
- 2p_\lambda k_\rho - 2p_\rho k_\lambda\right]
\nonumber\\ 
Y_{\lambda\rho}(\Omega,{\cal Q}) & = &
\int\frac{d^4p}{(2\pi)^3}\delta(p^2)\eta_e(p)
\left[
\frac{j_{\lambda\rho}}{q^2 - 2p\cdot q} + 
(q \rightarrow -q) \right]\,,
\end{eqnarray}
with
\begin{eqnarray}
\label{defj}
j_{\lambda\rho} = 
\frac{1}{4}(2p - q)_\lambda [(2k\cdot p - q\cdot k)p_\rho - 
(k\cdot p)q_\rho 
+ (p\cdot q)k_\rho] + (\lambda \leftrightarrow \rho) \,.
\end{eqnarray}
The terms $X^{(1,2)}_{\lambda\rho}$ can be expressed very simply in
terms of integrals over the electron distribution functions while
$Y_{\lambda\rho}(0,{\cal Q}\rightarrow 0)$ can be evaluated following
the techniques used to evaluate similar integrals in Ref.\
\cite{gravnu}. However, for our purposes here, the explicit results
for them are not needed.  For consistency, we just note that, despite
appearances, $(2v^\lambda v^\rho - \eta^{\lambda\rho})
Y_{\lambda\rho}(0,{\cal Q}\rightarrow 0)$ is well-defined (see 
Appendix~\ref{app:gaugeinv}).  Since each expression in Eq.\
(\ref{Bxiresults}) contains an explicit factor of $k^2$,we then obtain
\begin{eqnarray}
\label{gaugeinvproof}
B_\xi(\pm K,K) = 0 \,,
\end{eqnarray}
which completes the proof of Eq.\ (\ref{disprelformula}).

\subsection{Calculation of $B^{(4)}$}\label{subsec:B4}
As a byproduct of the gauge invariance proof given above, it follows
that the longitudinal part of $\Delta^{(4)}_{W\mu\nu}$ does not
contribute to the dispersion relations, which involve
$B^{(4)}(\pm K,K)$. Therefore, to obtain the dispersion relations, we
can use
\begin{eqnarray}
k^2 = \omega^2 - K^2 = 0 
\label{ksq=0}
\end{eqnarray}
and substitute
\begin{eqnarray}
\label{Wpropreplace}
\Delta^{(4)}_{W\mu\nu}(q) \rightarrow \eta_{\mu\nu}\frac{q^2}{M_W^4}
\end{eqnarray}
in the formulas for the $G^{(X)}_{\lambda\rho}$.  Corresponding to
each $G^{(X)}_{\lambda\rho}$, there is a term in $B^{(4)}$, and
according to Eq.\ (\ref{AB4}) we have
\begin{equation}
\label{B4formula}
B^{(4)}(K,K) = \phi^{\rm ext}\left(
\sum_{X=A,B,C,D}G^{(X)}(0,\calQ\rightarrow 0)\right)\,,
\end{equation}
where
\begin{eqnarray}
\label{GXscalar}
G^{(X)}(\Omega,\calQ) = (2v^\lambda v^\rho - \eta^{\lambda\rho})
\frac{1}{2}\mbox{Tr}k\llap / G^{(X)}_{\lambda\rho}\,.
\end{eqnarray}
We consider the computation of each of the $G^{(X)}(\Omega,\calQ)$
separately.

\subsubsection{Computation of $G^{(A)}$}
In the term containing $\eta_e(p^\prime)$ in the integral given in
Eq.\ (\ref{GA}), we make the change of variables $p \rightarrow p +
q$.  Then using the replacement indicated in Eq.\ (\ref{Wpropreplace})
for the $W$ propagator, and remembering that we are setting
$m_e\rightarrow 0$, we obtain
\begin{eqnarray}
\label{GAaux}
\frac{1}{2}\mbox{Tr}k\llap /G^{(A)}_{\lambda\rho} = 
-\, \frac{g^2}{2M_W^4} \int\frac{d^4p}{(2\pi)^3}
\delta(p^2)\eta_e(p)\Bigg[
\frac{(p - k)^2}{q^2 - 2p\cdot q} \Big(
\mathscr M_{\lambda\rho}(q) - \mathscr M'_{\lambda\rho}(q) \Big) \quad
\nonumber\\
+ \frac{(p - k^\prime)^2}{q^2 + 2p\cdot q}
\Big(\mathscr M_{\lambda\rho}(-q) + \mathscr M'_{\lambda\rho}(-q)\Big) 
\Bigg] \,,
\end{eqnarray}
where 
\begin{eqnarray}
\mathscr M_{\lambda\rho}(q) 
&=& -\frac12 \mbox{Tr} \Big[
k\llap / 
p\llap / V_{\lambda\rho}(p,p - q) (p\llap / - q\llap /) \Big] 
\,, \label{M}\\
\mathscr M'_{\lambda\rho}(q) &\equiv& -\frac12 \mbox{Tr} \Big[
k\llap / 
p\llap / V_{\lambda\rho}(p,p - q) (p\llap / - q\llap /) \gamma_5 \Big]
\,.
\label{M'}
\end{eqnarray}
In writing these forms, we have used the cyclicity of trace and the
contraction identity $\gamma_\alpha \rlap/k \gamma^\alpha =
-2\rlap/k$. We have also used the identity $C\gamma_\mu C^{-1} =
-\gamma^T_\mu$ to invert the order of the gamma matrices inside the
trace in the second term in Eq.\ (\ref{GAaux}), as well as the
property $V_{\lambda\rho}(p,p') = V_{\lambda\rho}(p',p)$.  The trace
operation can be carried out straightforwardly. The 
$\mathscr M'_{\lambda\rho}$ terms does not contribute to the final result. As
for the $\mathscr M_{\lambda\rho}$ terms, we carry out the integral
over $p^0$ with the help of the $\delta$-function and substitute the
resulting expression in Eq.\ (\ref{GXscalar}). This gives
\begin{eqnarray}
\label{GAintegral}
G^{(A)}(\Omega,\vec\calQ) & = & 
-\, \frac{g^2}{M_W^4} \int\frac{d^3p}
{(2\pi)^3 2E}\Bigg[
\frac{k^2-2k\cdot p}{q^2 - 2p\cdot q}
\mathscr M(p, q) f_e 
+ \frac{k^2 + 2k\cdot p}{q^2 + 2p\cdot q}
\mathscr M(-p,q) f_{\bar e}\nonumber\\ 
& & + 
\frac{k'^2 - 2k'\cdot p}{q^2 + 2p\cdot q}
\mathscr M(p,-q) f_e
+ \frac{k'^2 + 2k'\cdot p}{q^2 - 2p\cdot q}
\mathscr M(-p,-q) f_{\bar e} \nonumber\\ 
& & - \frac12
k\cdot p\left[k^2 + k'^2 - 2p\cdot(k + k')\right]f_e
+ \frac12 k\cdot p\left[k^2 + k'^2 + 2p\cdot(k + k')\right]f_{\bar e} 
\Bigg] \,, \nonumber\\* 
\end{eqnarray}
where
\begin{eqnarray}
\label{MAuu}
\mathscr M(p,q) & = & \Big[ -4(p\cdot v)^2 + 4 p\cdot v \, q\cdot v
- (q\cdot v)^2 \Big] k\cdot p \nonumber\\* 
&& + 
\Big[ -2p\cdot v  + q\cdot v \Big] k\cdot v \, p\cdot q +
\Big[ 2p\cdot v - q\cdot v \Big] p\cdot v \, k\cdot q \,.
\end{eqnarray}

To evaluate $G^{(A)}(0,\vec\calQ\rightarrow 0)$ we
proceed as follows.  We carry out the integration in the rest frame of
the background. In that frame Eqs.\ (\ref{vatrest}) and (\ref{kqatrest}) hold,
and similarly we write
\begin{equation}
\label{patrest}
p^\mu = (\calE,\vec\calP) \,,
\end{equation}
where $\calE=|\vec\calP|$ because of the $\delta$-function that
appears in Eq.\ (\ref{GAaux}).  In addition we use the fact that
$k^\mu$ satisfies Eq.\ (\ref{ksq=0}).  We next evaluate Eq.\
(\ref{GAintegral}) in the static limit $\Omega = 0$ and take
$\vec\calQ \rightarrow 0$ afterwards. The terms without any
denominator can be easily evaluated in this limit by putting
$k=k'$. The results can be conveniently expressed by defining the
integrals
\begin{eqnarray}
I_{\mu_1\mu_2\cdots\mu_r} \equiv \int {d^3\calP\over
(2\pi)^3 2\calE} \; \calP_{\mu_1} \calP_{\mu_2} \cdots \calP_{\mu_r}
\Big[ f_e(\calE) + (-1)^r f_{\bar e} (\calE) \Big] \,,
\label{I}
\end{eqnarray}
so that the contribution of these terms to
$G^{(A)}(0,\vec\calQ\rightarrow 0)$ can be written as
\begin{eqnarray}
-\, \left(\frac{g^2}{M_W^4}\right) 2 k^\mu k^\nu I_{\mu\nu} \,.
\label{GAregulars}
\end{eqnarray}
It is now useful to introduce the scalar integrals, for any fermion
$f$,
\begin{eqnarray}
J^{(f)}_n &\equiv& \int {d^3\calP\over
(2\pi)^3 2\calE} \calE^n 
\Big[ f_f(\calE) + (-1)^n f_{\bar f} (\calE) \Big] \nonumber\\*
&=& {1\over 4\pi^2}\int d\calE \; \calE^{n+1} 
\Big[ f_f(\calE) + (-1)^n f_{\bar f} (\calE) \Big] \,,
\label{Jn}
\end{eqnarray}
for which we give explicit formulas in Sec.~\ref{subsec:disp}.
Further simplification is then obtained by noting that the integrals in
Eq.\ (\ref{I}) can be determined in terms of the $J^{(e)}_n$; e.g.,
\begin{eqnarray}
I_\mu & = & J^{(e)}_1 v_\mu \,, \nonumber\\*
I_{\mu\nu} & = & \frac{1}{3} 
(-\eta_{\mu\nu} + 4 v_\mu v_\nu) J^{(e)}_2 \,, \nonumber\\*
I_{\mu\nu\lambda} &= & \left[ -\frac13 \Big(\eta_{\mu\nu} v_\lambda +
\eta_{\nu\lambda} v_\mu + \eta_{\lambda\mu} v_\nu \Big) + 2v_\mu v_\nu
v_\lambda \right] J^{(e)}_3 \,. 
\label{Ieval}
\end{eqnarray}
Thus the contribution of Eq.\ (\ref{GAregulars}) reduces to
\begin{eqnarray}
-\, \left(\frac{g^2}{M_W^4}\right)
\frac{8}{3} \omega^2 J^{(e)}_2 \,.
\label{GAregular}
\end{eqnarray}

Despite appearances, the remaining terms in Eq.\ (\ref{GAintegral}),
which contain $q$ in the denominator, yield a well-defined result in
the $(\Omega = 0,\vec\calQ\rightarrow 0)$ limit.  We relegate the
details of the calculation to Appendix~\ref{appGA} and give the final
result here,
\begin{equation}
\label{GAfinal}
G^{(A)}(0,\vec\calQ\rightarrow 0) =
\frac{g^2}{M_W^4} \Bigg[
\frac{28}{3}\omega^2 J^{(e)}_2 + 4\omega J^{(e)}_3 \Bigg] \,,
\end{equation}
including the contribution of the regular terms given in 
Eq.\ (\ref{GAregular}).

\subsubsection{Computation of $G^{(B)}$}
For $G^{(B)}$, we can directly put $q = 0$ and use Eq.\
(\ref{Wpropreplace}) in Eq.\ (\ref{GHB}).  We then calculate
\begin{eqnarray}
\frac{1}{2}\mbox{Tr}k\llap /G^{(B)}_{\lambda\rho}(0,0) & = &
\frac{g^2}{2M_W^4}\int\frac{d^4p}{(2\pi)^3}\delta(p^2)\eta_e(p)
\Bigg\{
k^2\left[k\cdot p\, \eta_{\lambda\rho} - 4k_\lambda p_\rho -
4k_\rho p_\lambda + 4p_\lambda p_\rho\right] \nonumber\\*
& & \null + 4 k\cdot p \left[
2k_\lambda p_\rho + 2k_\rho p_\lambda - p_\lambda p_\rho - k_\lambda
k_\rho \right] \Bigg\} \,.
\end{eqnarray}
Using Eq.\ (\ref{ksq=0}) now and carrying out the integral over $p^0$,
from Eq.\ (\ref{GXscalar}) we obtain
\begin{eqnarray}
G^{(B)}(0,0) & = & \frac{g^2}{M_W^4}\int\frac{d^3 p}{(2\pi)^3 2E}
\Bigg\{ 
8k\cdot p \Big[2k\cdot v\, p\cdot v - k\cdot p \Big]
(f_e + f_{\bar e}) \nonumber\\*
& & \null - 4k\cdot p \Big[(p\cdot v)^2 + (k\cdot v)^2 \Big]
(f_e - f_{\bar e}) \Bigg\} \,.
\end{eqnarray}
This can be written as
\begin{eqnarray}
\label{GBprelim}
G^{(B)}(0,0) & = & \frac{g^2}{M_W^4}\Bigg\{
16(k\cdot v)k^\mu v^\nu I_{\mu\nu} - 8 k^\mu k^\nu I_{\mu\nu}
- 4 k^\mu v^\nu v^\lambda I_{\mu\nu\lambda} - 4(k\cdot v)^2k^\mu
I_\mu\Bigg\} \,,
\end{eqnarray}
and using Eq.\ (\ref{Ieval})
\begin{equation}
\label{GBfinal}
G^{(B)}(0,0) = \frac{g^2}{M_W^4}\left\{
- 4\omega^3 J^{(e)}_1 + \frac{16}{3}\omega^2 J^{(e)}_2 - 
4\omega J^{(e)}_3\right\} \,.
\end{equation}
%

\subsubsection{Computation of $G^{(C)}$ and $G^{(D)}$}
Proceeding in similar fashion, we get
\begin{equation}
\frac{1}{2}\mbox{Tr}k\llap /G^{(C)}_{\lambda\rho}(0,0) =
\frac{g^2}{2M_W^4}\int\frac{d^4p}{(2\pi)^3}\delta(p^2)\eta_e(p)
(p - k)^2[\eta_{\lambda\rho}(k\cdot p) + 
p_\lambda k_\rho + k_\lambda p_\rho] \,.
\end{equation}
Using Eq.\ (\ref{ksq=0}) and Eq.\ (\ref{GXscalar}), we can now write
\begin{eqnarray}
G^{(C)}(0,0) 
& = &
- \, \frac{g^2}{M_W^4} \int\frac{d^3p}{(2\pi)^3 2E} \;
4 k\cdot p 
\Big[p\cdot v\, k\cdot v - k\cdot p \Big] (f_e + f_{\bar e}) \nonumber\\
& = & -\, \frac{g^2}{M_W^4} \bigg\{4(k\cdot v)k^\mu v^\nu
I_{\mu\nu} - 4k^\mu k^\nu I_{\mu\nu} 
\bigg\} \,,
\end{eqnarray}
and similarly it can be verified that $G^{(D)}(0,0)$ is given
by the same expression.  Using Eq.\ (\ref{Ieval}) then
\begin{equation}
\label{GCDfinal}
G^{(C)}(0,0) = G^{(D)}(0,0) = \left(\frac{g^2}{M_W^4}\right)
\frac{4}{3}\omega^2 J^{(e)}_2  \,.
\end{equation}
%

%
%
\section{Neutrino dispersion relation}\label{subsec:disp}
\setcounter{equation}{0}
Collecting the results for the $G^{(X)}(0,\calQ\rightarrow 0)$
given in Eqs.\ (\ref{GAfinal}), (\ref{GBfinal}) and (\ref{GCDfinal}) 
and substituting them in Eq.\ (\ref{B4formula}), we finally obtain
\begin{equation}
\label{B4final}
B^{(4)}(\omega,K) = {g^2 \over M_W^4} \Bigg[ - 4\omega^3 J^{(e)}_1 
+ \frac{52}{3} \omega^2 J^{(e)}_2 \Bigg] \phi^{\rm ext}
\end{equation}
for $\omega=\pm K$, which is the result that must be substituted
in Eq.\ (\ref{disprelformula}) for the dispersion
relation of the electron neutrino. For $\nu_{\mu,\tau}$, 
Eq.\ (\ref{disprelformula}) holds but the $B^{(4)}$ term is absent.

The integrals defined in Eq.\ (\ref{Jn}) cannot be evaluated in a
general way for arbitrary distributions and $m_f \not = 0$, but for
$m_e = 0$ they can be performed exactly to yield the following result
in terms of the chemical potential and temperature of the electron
gas:
\begin{eqnarray}
J_1^{(e)} &=& {\mu_e^3 \over 12\pi^2} + {\mu_eT^2 \over 12}\,,
\nonumber\\* 
J_2^{(e)} &=& {\mu_e^4\over 16\pi^2} + {\mu_e^2T^2 \over 8} + {7\pi^2T^4
\over 240} \,.
\label{J12massless}
\end{eqnarray}
For our present purposes, it is sufficient to note that, in any case,
$J^{(f)}_1$ and $J^{(f)}_2$ are simply related to the number density
$n_f$ and energy density $\rho_f$ of the particles in the background
medium by the formulas
\begin{eqnarray}
\label{J12}
J^{(f)}_1 &=& \frac14 (n_f - n_{\bar f}) \,, \nonumber\\*
J^{(f)}_2 &=& \frac14 (\rho_f + \rho_{\bar f}) \,.
\end{eqnarray}
%
%
%
Thus, using Eq.\ (\ref{B4final}), the dispersion relation becomes
\begin{eqnarray}
\label{disprelfinal}
\omega_K & = & K + 2K\phi^{\rm ext}(1 + \gamma) + 
(1 + \phi^{\rm ext})b_{\submat} + b^{(2)}_G \nonumber\\
\bar\omega_K & = & K + 2K\phi^{\rm ext}(1 + \bar\gamma) + 
(1 + \phi^{\rm ext}){\bar b}_{\submat} - b^{(2)}_G \,,
\end{eqnarray}
where
\begin{eqnarray}
\label{gammafactor}
\gamma & = & \left(\frac{g^2}{4M^4_W}\right)\left[-2K(n_e - n_{\bar e})
+ \frac{26}{3}(\rho_e + \rho_{\bar e})\right] \qquad  \mbox{for $\nu_e$}
\nonumber\\
\bar \gamma & = & \left(\frac{g^2}{4M^4_W}\right)\left[
2K(n_e - n_{\bar e}) + \frac{26}{3}(\rho_e + \rho_{\bar e})\right]\phantom{-}
\qquad \mbox{for $\bar\nu_e$}\,,
\end{eqnarray}
with $\gamma$ ($\bar\gamma$) being zero for 
$\nu_{\mu,\tau}$ ($\bar\nu_{\mu,\tau}$).

The other terms in Eq.\ (\ref{disprelfinal}) are the following.
In Ref.~\cite{gravnu}, we calculated the contributions
to $b^{(2)}_G$ for various kinds of
background particles --- non-relativistic or ultra-relativistic,
non-degenerate or degenerate, arriving at the final formula
given in Eq.\ (4.30) there. For the electrons,
we use here the results for the
ultra-relativistic case since it is the one that corresponds
to the limit $m_e = 0$ that we have adopted in the present paper. Then, 
\begin{eqnarray}
\label{bG2}
b^{(2)}_G & = & \left(\frac{g^2}{4M^2_W}\right)\phi^{\rm ext}
\times
\left\{ \begin{array}{ll}
\calJ_e +
\sum_f X_f \calJ_f  & \mbox{for $\nu_e$}\,,\\
&\\
\sum_f X_f \calJ_f   & \mbox{for
$\nu_\mu,\nu_\tau$} 
\end{array}\right.  
\end{eqnarray}
for $f = e,n,p$, where $X_f$ is the vector neutral-current coupling of
the fermion, and
\begin{eqnarray}
\calJ_e & = & -5(n_e - n_{\bar e}) \nonumber\\[12pt]
\calJ_N & = &
\left\{
\begin{array}{ll}
- \beta m_N n_N & \mbox{classical nucleon gas}\\
&\\
- \, \frac{\textstyle 3n_N}{\textstyle v_{FN}^2} & 
\mbox{degenerate nucleon gas\,,}
\end{array}\right. 
\end{eqnarray}
for $N = n,p$,
where $v_{FN}$ stands for the Fermi velocity of the nucleon gas.

The term $b_{\submat}$ can be split into two parts.  One
is the usual Wolfenstein term, which is given by
\begin{eqnarray}
\label{bmat2}
b^{(2)}_{\submat} & = & \left(\frac{g^2}{4M^2_W}\right)
\times
\left\{ \begin{array}{ll}
(n_e - n_{\bar e}) +
\sum_f X_f(n_f - n_{\bar f})   & \mbox{for $\nu_e$}\,,\\
&\\
\sum_f X_f(n_e - n_{\bar e})   & \mbox{for
$\nu_\mu,\nu_\tau$} 
\end{array}\right.  
\end{eqnarray}
for the neutrinos, and ${\bar b}^{(2)}_{\submat} = -b^{(2)}_{\submat}$ for the
antineutrinos. The second part consists of the corrections
of $O(1/M^4_W)$. Using the results of Ref.\ \cite{dnt}
[in particular Eq.\ (3.27)], but putting $m_e\rightarrow 0$,
and neglecting the terms that arise if there are neutrinos
and other charged leptons in the background, 
we can write this part in the form
\begin{eqnarray}
\label{bmat4}
b^{(4)}_{\submat} & = & \left(\frac{g^2}{4M^4_W}\right)
\times
\left\{ \begin{array}{ll}
-\frac{8}{3}K(\rho_e + \rho_{\bar e}) & \mbox{for $\nu_e$} \,,\\
&\\
0   & \mbox{for
$\nu_\mu,\nu_\tau$} 
\end{array}\right.  
\end{eqnarray}
with ${\bar b}^{(4)}_{\submat} = b^{(4)}_{\submat}$ for the antineutrinos.

\section{Discussions and Conclusions}
\label{sec:conclusions}
\setcounter{equation}{0}

As the formulas given in Eqs.\ (\ref{disprelfinal}) and (\ref{gammafactor})
show, the effects of the induced gravitational terms on the neutrino
dispersion relation are more noticeable at high energies.
An important point to remember is that the calculations that we
have presented rely on the approximation made in 
Eq.\ (\ref{propagatorapprox}) for the $W$-boson propagator. This implies
that the results are valid as long as $(p - k)^2 \ll M^2_W$,
which translates to $K\langle E_e\rangle \ll M^2_W$, where
$\langle E_e\rangle$ denotes a typical average energy of the electrons
in the background. For backgrounds with a temperature of the order
of the electron mass or somewhat above, this implies that
$K \ll 10^7 GeV$. Thus, if we consider values of $K \approx 10^5 - 10^6 GeV$,
for which our results hold, we find that the induced gravitational
effects can be substantial if the gravitational potential is not too
small. In fact, from Eq.\ (\ref{gammafactor}), it follows in that case
that
\begin{equation}
\label{lastterm}
2K\gamma\phi^{\rm ext} 
\approx -\left(\frac{g^2}{M^2_W}\right)(n_e - n_{\bar e})
\left(\frac{K}{10^2 GeV}\right)^2\phi^{\rm ext} 
\end{equation}
for $\nu_e$, with $\bar\gamma = - \gamma$ for $\bar\nu_e$,
and zero for the other neutrino and antineutrino types. 
Moreover, it is possible that,
for a certain range of values of the neutrino momentum,
the term in Eq.\ (\ref{lastterm}) is the dominant one in the
neutrino dispersion relation.
These considerations may be applicable, for example, in the environments 
of Active Galactic Nuclei. It has been observed that
in such systems, high energy neutrinos
can have resonant spin-flavor transitions due to the combined
effects of the gravitational interactions and the presence
of large magnetic fields\cite{wudkaetal}. Since in such environments
$\phi^{\rm ext}\approx 1/10 - 1/100$
in the region in which the neutrinos are produced,
the momentum-dependent gravitational terms modify
the neutrino dispersion relation in a substantial way,
and therefore they can have
observable consequences in the determination of the high
energy neutrino fluxes from such systems. 
The calculations
that we have presented, and our result for the neutrino
dispersion relation, are useful in this and possibly other contexts, and 
provide a firm basis to study their possible effects in detail.

\appendix
%
%
\section*{Appendices}
\section{Proof of Eq. (\ref{gaugeinvproof})}\label{app:gaugeinv}
\setcounter{equation}{0}
To prove the statement that 
$(2v^\lambda v^\rho - \eta^{\lambda\rho})
Y_{\lambda\rho}(0,{\cal Q}\rightarrow 0)$
is well defined, we consider the integral formula
for $Y_{\lambda\rho}(\Omega,{\cal Q})$
given in Eq.\ (\ref{I12J}). 
Carrying out the integration over $p^0$, in the rest frame of medium, 
and using the fact that, considered as a function of $p$ and $q$,
\begin{eqnarray}
j_{\lambda\rho}(-p,q) = -j_{\lambda\rho}(p,-q) \,,
\end{eqnarray}
we have
\begin{eqnarray}
Y_{\lambda\rho}(\Omega,\calQ) \equiv \
\int\frac{d^3P}{(2\pi)^3 2\calE}(f_e - f_{\bar e})
\left[
\frac{j_{\lambda\rho}}{q^2 - 2p\cdot q} + 
(q \rightarrow -q) \right ] \,,
\end{eqnarray}
where now
\begin{eqnarray}
p^\mu = (\calE,\vec\calP) \qquad \calE = |\vec\calP| \,.
\end{eqnarray}
We consider separately the functions
\begin{eqnarray}
\label{Y12}
Y_\eta & = & \eta^{\lambda\rho} Y_{\lambda\rho}\nonumber\\
Y_{vv} & = & v^\lambda v^\rho Y_{\lambda\rho} \,,
\end{eqnarray}
and show explicitly that each one has a well defined value
in the limit $(\Omega = 0, \vec\calQ\rightarrow 0)$,
which proves the statement.

{}From Eq.\ (\ref{defj}) we obtain 
\begin{eqnarray}
\eta^{\lambda\rho}j_{\lambda\rho} = 
-(k\cdot p)\frac{1}{2}(2p\cdot q - q^2) 
\end{eqnarray}
which implies that
\begin{eqnarray}
Y_\eta & = & -\int\frac{d^3 P}{(2\pi)^3 2\calE}
(f_e - f_{\bar e})k\cdot p \nonumber\\
& = & \frac{1}{4}(n_e - n_{\bar e})k\cdot v \,.
\end{eqnarray}
For $Y_{vv}$, setting $\Omega = 0$ we obtain
\begin{eqnarray}
Y_{vv}(0,\vec\calQ) & = & \frac{1}{4}\int\frac{d^3P}{(2\pi)^3}(f_e -
f_{\bar e}) 
(2\calE^2\omega - 2\calE\vec\calP\cdot\vec K)\left[
\frac{1}{2\vec\calP\cdot\vec\calQ - \calQ^2} -
\frac{1}{2\vec\calP\cdot\vec\calQ + \calQ^2}
\right] \nonumber\\
& & \mbox{} +
\frac{1}{4}\int\frac{d^3\calP}{(2\pi)^3}(f_e - f_{\bar e})
(\calE\vec\calQ\cdot\vec K - \omega\vec\calP\cdot\vec\calQ)\left[
\frac{1}{2\vec\calP\cdot\vec\calQ - \calQ^2} +
\frac{1}{2\vec\calP\cdot\vec\calQ + \calQ^2}
\right] \,.
\end{eqnarray}
The integrals with the factors $2\calE\vec\calP\cdot\vec K$ and 
$\calE\vec\calQ\cdot\vec K$ are zero by symmetric integration
since the corresponding integrands are odd under
$\vec\calP \rightarrow -\vec\calP$.
The remaining two integrals are evaluated by making the change
of variables $\vec\calP\rightarrow \vec\calP \pm \frac{1}{2}\vec\calQ$
when the denominator is $2\vec\calP\cdot\vec\calQ \mp \calQ^2$,
respectively, and then taking the limit $\vec\calQ \rightarrow 0$.
In this way,
\begin{eqnarray}
Y_{vv}(0,\vec\calQ\rightarrow 0) & =  &
\frac{1}{4}\int\frac{d^3\calP}{(2\pi)^3 2\calE}\frac{d}{d\calE}
[2\omega \calE^2(f_e - f_{\bar e})]
-
\frac{1}{4}\int\frac{d^3\calP}{(2\pi)^3}[\omega(f_e - f_{\bar e})]\nonumber\\
& = & \frac{1}{4}\int\frac{d^3\calP}{(2\pi)^3}\omega\frac{d}{d\calE}[\calE(f_e
- f_{\bar e})] \,.
\end{eqnarray}
%

%
%
\section{The limit $G^{(A)}(\Omega=0,\vec\mathcal Q\to 0)$}\label{appGA}
\setcounter{equation}{0}
The expression for $G^{(A)}$ appears in Eq.\ (\ref{GAintegral}). 
In the text, we have evaluated the terms without any denominator. Let
us denote the remaining terms by $\widehat G^{(A)}$. For
these terms, we first put $\Omega = 0$. The resulting expression for
$\mathscr M(p,q)$ is written as
\begin{eqnarray}
\mathscr M_0(p, \vec \calQ) = 
- 4\calE^3\omega + 4\calE^2\vec\calP\cdot\vec \calK +
2\calE\omega\vec\calP\cdot\vec\calQ 
- 2\calE^2\vec\calK\cdot\vec\calQ \,,
\end{eqnarray}
which follows directly from the definition in Eq.\ (\ref{MAuu}). We
can then write these potentially singular terms in the form
\begin{eqnarray}
\widehat G^{(A)}(0,\vec\mathcal Q) & = & 
-\, \frac{g^2}{M_W^4} \int\frac{d^3\calP}
{(2\pi)^3 2\calE}\Bigg[ 
\frac{ (k^2 - 2k\cdot p) \mathscr M_0(p,\vec\mathcal Q)}
{2\vec\calP \cdot \vec\calQ - \calQ^2} \; f_e(\calE) 
- \frac{(k^2 + 2k\cdot p) \mathscr M_0(-p,\vec\mathcal
Q)}{2\vec\calP \cdot \vec\calQ + \calQ^2} 
f_{\bar e}(\calE)\nonumber\\ 
&& \qquad - \frac{ (k'^2 - 2k'\cdot p) 
\mathscr M_0(p,-\vec\mathcal Q)}
{2\vec\calP \cdot \vec\calQ + \calQ^2} \; f_e(\calE) 
+ \frac{(k'^2 + 2k'\cdot p) \mathscr M_0(-p,-\vec\mathcal
Q)}{2\vec\calP \cdot \vec\calQ - \calQ^2} 
f_{\bar e}(\calE) \Bigg] \nonumber\\ 
&=& -\, \frac{g^2}{M_W^4} 
\Big[ \widehat G_1^{(A)}(0,\vec\mathcal Q) + \widehat
G_2^{(A)}(0,\vec\mathcal Q) \Big] \,,
\label{If}
\end{eqnarray}
where $\widehat G_1$ comes from the first pair of terms and $\widehat
G_2$ from the second pair.

The integrals in this equation can be conveniently expressed in terms
of the following ones: 
\begin{eqnarray}
\calF_{n,m} & = & \int\frac{d^3\calP}{(2\pi)^3 2\calE }
\; \calE^n (\vec\calP\cdot\vec\calQ)^m \Bigg[ 
\frac{f_e(\calE)} {2\vec\calP\cdot\vec\calQ - \calQ^2} - (-1)^{n+m}
\frac{f_{\bar e}(\calE)} {2\vec\calP\cdot\vec\calQ + \calQ^2} \Bigg] \,,
\end{eqnarray}
as well as
\begin{eqnarray}
\calG^{(n)}_{i_1 i_2 \cdots i_r} & = &
\int\frac{d^3\calP}{(2\pi)^3 2\calE } 
\calE^n \calP_{i_1}\calP_{i_2} \cdots \calP_{i_r} \Bigg[  
\frac{f_e(\calE)}{2\vec\calP\cdot\vec\calQ - \calQ^2} - (-1)^{n+r}
\frac{f_{\bar e}(\calE)} {2\vec\calP\cdot\vec\calQ + \calQ^2} \Bigg]
\,,
\label{G}
\end{eqnarray}
where $i_1$ etc are spatial indices. Using these notations, we can
write the first pair of terms for $\widehat G^{(A)}$ as
\begin{eqnarray}
\label{term1}
\widehat G_1^{(A)}(0,\vec\mathcal Q)
&=& 
16\omega^2 \calF_{4,0} + 8\omega\vec
\calK\cdot\vec\calQ \calF_{3,0} - 8\omega^2 \calF_{2,1} 
\nonumber\\ & & 
- 32\omega \calK_i \calG_i^{(3)} 
- 8\vec \calK\cdot\vec\calQ\, \calK_i \calG_i^{(2)} 
+ 16 \calK_i \calK_j \calG_{ij}^{(2)}
+ 8\omega \calQ_i \calK_j \calG_{ij}^{(1)} 
\,,
\end{eqnarray}
where we have now used Eq.\ (\ref{ksq=0}).  To simplify further, we
first notice that by elementary arguments the integrals in Eq.\
(\ref{G}) can be expressed in terms of the integrals $\calF_{n,m}$:
\begin{eqnarray}
\calG_i^{(n)} & = & \frac{\calQ_i}{\calQ^2}F_{n,1} \nonumber\\
\calG^{(n)}_{ij} & = & \frac{1}{2} \Bigg[ \left(
F_{n + 2,0} - \frac{F_{n,2}}{\calQ^2} \right) \delta_{ij}
- \left( F_{n + 2,0} -
\frac{3F_{n,2}}{\calQ^2}  
\right) \frac{\calQ_i \calQ_j}{\calQ^2} \Bigg] \,.
\end{eqnarray}
The next step is to realize that the scalar integrals $\calF_{n,m}$
for a particular value of $m$ can be expressed in terms of
$\calF_{n,m}$ for smaller values of $m$ and the integrals $J^{(e)}_n$
defined in Eq.\ (\ref{Jn}). For example, 
\begin{eqnarray}
\calF_{n,1} &=& \frac12 \Big[ J^{(e)}_n + \calQ^2 \calF_{n,0} \Big] \,,
\nonumber\\* 
\calF_{n,2} &=& \frac12 \calQ^2 \calF_{n,1} \,,
\label{Fiterative}
\end{eqnarray}
Finally, in $\calF_{n,0}$, the angular integration can be performed
exactly to give
\begin{eqnarray}
\calF_{n,0} = {1\over 16\pi^2} \int_0^\infty d\calE \; {\calE^n \over
\calQ} \; \ln \left( {1-(\calQ/2\calE) \over 1+(\calQ/2\calE)} \right)
\times \Big[ f_e(\calE) + (-1)^n f_{\bar e} (\calE) \Big] \,.
\end{eqnarray}
For small $\calQ$, the logarithm can be expanded in power series and
we obtain
\begin{eqnarray}
\calF_{n,0} & = & -\, \frac{1}{4} \; J^{(e)}_{n-2} - 
\frac{\calQ^2}{48} \; J^{(e)}_{n - 4} + O\left(\calQ^4\right) \,.
\end{eqnarray}
The integrals for larger values of $m$ can be obtained by 
using Eq.\ (\ref{Fiterative}).

When we put the results of these integrals, Eq.\ (\ref{term1}) reduces 
to the form
\begin{eqnarray}
\widehat G_1^{(A)}(0,\vec\mathcal Q) &=& 
- 12\omega^2 J^{(e)}_2
+ \frac{4(\vec \calK\cdot\vec\calQ)^2}{\calQ^2}J^{(e)}_2
- \frac{16\omega(\vec \calK\cdot\vec\calQ)}{\calQ^2}J^{(e)}_3 + 
{\cal O}(\calQ^2)\,, 
\end{eqnarray}
In a similar fashion, the other pair of terms in $\widehat G^{(A)}$
can be evaluated. The result is
\begin{eqnarray}
\widehat G_2^{(A)}(0,\vec\mathcal Q) &=& 
- 12\omega^2J^{(e)}_2 -
\frac{4(\vec \calK\cdot\vec\calQ)^2}{\calQ^2}J^{(e)}_2
+ \frac{16\omega(\vec \calK\cdot\vec\calQ)}{\calQ^2}J^{(e)}_3 
- 8\omega J^{(e)}_3  + \mathcal O (\calQ^2)\,.
\end{eqnarray}
Summing these two contributions and reinstating the overall factor, we
can write
\begin{eqnarray}
\widehat G^{(A)} (0,0) = {g^2\over M_W^4} \bigg[ 
12 \omega^2 J^{(e)}_2 + 4\omega J^{(e)}_3 \bigg] \,.
\label{GAsingular}
\end{eqnarray}
Adding this with the contribution of the regular terms obtained in
Eq.\ (\ref{GAregular}), we get the total result shown in Eq.\
(\ref{GAfinal}).

%
%
An alternative evaluation of Eq.\ (\ref{GAsingular}) is as follows.
In the $\mathcal{Q}$-dependent integrals we make the change of
variables $\vec\calP \rightarrow \vec\calP \pm
\frac{1}{2} \vec\calQ$ in those terms containing in the denominator
the factors $\vec\calP\cdot\vec\calQ \mp \calQ^2$,
respectively. By expanding the integrands in powers of $\calQ$ and
taking the limit $\calQ\rightarrow 0$ we obtain in this fashion
\begin{eqnarray}
\widehat G^{(A)}(0,\vec\mathcal{Q}\rightarrow 0) &=&
-\, \frac{g^2}{M_W^4} \int\frac{d^3\calP}
{(2\pi)^3 2\calE } \, 2k\cdot p \Bigg\{
\bigg[ |\vec\calP - \vec \calK|^2 
\frac{d}{d\calE }\left(\calE  f_e(\calE )\right) - 
\frac{d}{d\calE }\left[(\calE  - \omega)^2
\calE f_e\right] \bigg] \nonumber\\*
&& \hspace{2cm} -
\bigg[ |\vec\calP + \vec \calK|^2 
\frac{d}{d\calE }\left(\calE  f_{\bar e}(\calE )\right)
- \frac{d}{d\calE }\left[(\calE  + \omega)^2 \calE 
f_{\bar e}\right] \bigg]
\Bigg \} \nonumber\\ 
&=& \frac{g^2}{M_W^4} \int\frac{d^3\calP}
{(2\pi)^3 2\calE } \, 4k\cdot p \Bigg\{
\bigg[ - k\cdot p\frac{d(\calE f_e)}{d\calE}
+ \calE(\calE - \omega)f_e \bigg] \nonumber\\*
&& \hspace{2cm} -
\bigg[ k\cdot p\frac{d(\calE f_{\bar
e})}{d\calE} 
+ \calE(\calE + \omega)f_{\bar e} \bigg]
\Bigg \} \,,
\end{eqnarray}
using Eq.\ (\ref{ksq=0}) in the last step.  Carrying out an
integration by parts and combining some terms, we get
\begin{eqnarray}
\widehat G^{(A)}(0,\vec\mathcal{Q}\rightarrow 0) & = &
\frac{g^2}{M_W^4} \int\frac{d^3\calP}
{(2\pi)^3 2\calE} \Bigg[ 
12(k\cdot p)^2(f_e + f_{\bar e}) 
+ 4 k\cdot p (\calE^2 - \calE\omega) (f_e - f_{\bar e})
\Bigg] \,,
\end{eqnarray}
which can be written in the form
\begin{equation}
\label{GA1}
G^{(A)}(0,\vec\mathcal{Q}\rightarrow 0) =
\frac{g^2}{M_W^4} \bigg[ 
12k^\mu k^\nu I_{\mu\nu} + 4 k^\mu v^\nu v^\lambda I_{\mu\nu\lambda}
- 4 \omega k^\mu v^\nu I_{\mu\nu}
\bigg] \,,
\end{equation}
in terms of the integrals defined in Eq.\ (\ref{I}).
Substituting into Eq.\ (\ref{GA1}) the formulas given in 
Eq.\ (\ref{Ieval}), the result given in Eq.\ (\ref{GAsingular}) 
is reproduced.

%
%

\end{document}